\documentclass[11pt]{article}

\usepackage[utf8]{inputenc}
\usepackage[T1]{fontenc}

\usepackage{geometry}
\usepackage[bottom]{footmisc}
\usepackage{todonotes}
\usepackage{xspace}

\usepackage{comment}

\usepackage{nicefrac}
\usepackage[noend]{algpseudocode}

\usepackage{amssymb}
\usepackage{amsmath}
\usepackage{amsthm}
\usepackage{dsfont}
\usepackage{footnote}

\usepackage[ruled,vlined,linesnumbered]{algorithm2e}

\usepackage{thmtools} 
\usepackage{thm-restate}
\usepackage{lipsum}
\usepackage[framemethod=tikz]{mdframed}

\newenvironment{wrapper}[1]
{
	\begin{center}
		\begin{minipage}{\linewidth}
			\begin{mdframed}[hidealllines=true, backgroundcolor=gray!20, leftmargin=0cm,innerleftmargin=0.4cm,innerrightmargin=0.4cm,innertopmargin=0.4cm,innerbottommargin=0.4cm,roundcorner=10pt]
				#1}
			{\end{mdframed}
		\end{minipage}
	\end{center}
}

\usepackage{xcolor}
\usepackage{nameref}
\definecolor{ForestGreen}{rgb}{0.1333,0.5451,0.1333}
\definecolor{DarkRed}{rgb}{0.65,0,0}
\definecolor{Red}{rgb}{1,0,0}
\usepackage[linktocpage=true,
pagebackref=true,colorlinks,
linkcolor=DarkRed,citecolor=ForestGreen,
bookmarks,bookmarksopen,bookmarksnumbered]
{hyperref}
\usepackage{cleveref}

\usepackage{thm-restate}
\usepackage[tableposition=bottom]{caption}
\usepackage[all]{hypcap}
\usepackage{subcaption}
\usepackage{framed}
\usepackage[framemethod=tikz]{mdframed}
\usepackage[skins]{tcolorbox}

\makeatletter
\g@addto@macro{\maketitle}{\@thanks}
\makeatother

\newtheorem{theorem}{Theorem}[section]

\newtheorem{lemma}[theorem]{Lemma}
\newtheorem{definition}[theorem]{Definition}

\newtheorem{claim}[theorem]{Claim}

\newtheorem{remark}[theorem]{Remark}

\newcommand{\eps}{\epsilon}%
\newcommand{\p}{\textsc{P}}%
\newcommand{\poly}{\mathrm{poly}}
\newcommand{\polylog}{\mathrm{polylog}}%

\renewcommand{\algorithmiccomment}[1]{\bgroup\hfill$\rhd$~#1\egroup}

\newcounter{note}[section]

\newcommand{\dH}{\deg_H}

\renewcommand{\paragraph}[1]{\medskip\noindent\textbf{#1}}

\usepackage{etoolbox}
\usepackage{tikz}
\usetikzlibrary{tikzmark}
\usetikzlibrary{calc}

\errorcontextlines\maxdimen

\newcommand{\ALGtikzmarkcolor}{black}
\newcommand{\ALGtikzmarkextraindent}{4pt}
\newcommand{\ALGtikzmarkverticaloffsetstart}{-.5ex}
\newcommand{\ALGtikzmarkverticaloffsetend}{-.5ex}
\makeatletter
\newcounter{ALG@tikzmark@tempcnta}

\newcommand\ALG@tikzmark@start{%
	\global\let\ALG@tikzmark@last\ALG@tikzmark@starttext%
	\expandafter\edef\csname ALG@tikzmark@\theALG@nested\endcsname{\theALG@tikzmark@tempcnta}%
	\tikzmark{ALG@tikzmark@start@\csname ALG@tikzmark@\theALG@nested\endcsname}%
	\addtocounter{ALG@tikzmark@tempcnta}{1}%
}

\def\ALG@tikzmark@starttext{start}
\newcommand\ALG@tikzmark@end{%
	\ifx\ALG@tikzmark@last\ALG@tikzmark@starttext
	\else
	\tikzmark{ALG@tikzmark@end@\csname ALG@tikzmark@\theALG@nested\endcsname}%
	\tikz[overlay,remember picture] \draw[\ALGtikzmarkcolor] let \p{S}=($(pic cs:ALG@tikzmark@start@\csname ALG@tikzmark@\theALG@nested\endcsname)+(\ALGtikzmarkextraindent,\ALGtikzmarkverticaloffsetstart)$), \p{E}=($(pic cs:ALG@tikzmark@end@\csname ALG@tikzmark@\theALG@nested\endcsname)+(\ALGtikzmarkextraindent,\ALGtikzmarkverticaloffsetend)$) in (\x{S},\y{S})--(\x{S},\y{E});%
	\fi
	\gdef\ALG@tikzmark@last{end}%
}

\apptocmd{\ALG@beginblock}{\ALG@tikzmark@start}{}{\errmessage{failed to patch}}
\pretocmd{\ALG@endblock}{\ALG@tikzmark@end}{}{\errmessage{failed to patch}}
\makeatother

\title{Incremental (1-eps)-approximate dynamic matching in O(poly(1/eps)) update time}

\author{Joakim Blikstad\thanks{KTH Royal Institute of Technology, Sweden, Supported by the Swedish Research
Council (Reg. No. 2019-05622)} \thanks{Work done while visiting the Max Planck Institute for Informatics, Saarbrucken, Germany} \and Peter Kiss \thanks{University of Warwick} 
\footnotemark[2]}

\date{\today}

\begin{document}

\maketitle

\begin{abstract}

    In the dynamic approximate maximum bipartite matching problem we are given bipartite graph $G$ undergoing updates and our goal is to maintain a matching of $G$ which is large compared the maximum matching size $\mu(G)$. We define a dynamic matching algorithm to be $\alpha$ (respectively $(\alpha, \beta)$)-approximate if it maintains matching $M$ such that at all times $|M | \geq \mu(G) \cdot \alpha$ (respectively $|M|  \geq \mu(G) \cdot \alpha - \beta$). 

    We present the first deterministic $(1-\eps)$-approximate dynamic matching algorithm with $O(\poly(\eps^{-1}))$ amortized update time for graphs undergoing edge insertions. Previous solutions either required super-constant [Gupta FSTTCS'14, Bhattacharya-Kiss-Saranurak SODA'23] or exponential in $1/\eps$ [Grandoni-Leonardi-Sankowski-Schwiegelshohn-Solomon SODA'19] update time. Our implementation is arguably simpler than the mentioned algorithms and its description is self contained. Moreover, we show that if we allow for additive $(1, \eps \cdot n)$-approximation our algorithm seamlessly extends to also handle vertex deletions, on top of edge insertions. This makes our algorithm one of the few small update time algorithms for $(1-\eps)$-approximate dynamic matching allowing for updates both increasing and decreasing the maximum matching size of $G$ in a fully dynamic manner. 

    Our algorithm relies on the weighted variant of the celebrated Edge-Degree-Constrained-Subgraph (EDCS) datastructure introduced by [Bernstein-Stein ICALP'15]. As far as we are aware we introduce the first application of the weighted-EDCS for arbitrarily dense graphs. We also present a significantly simplified proof for the approximation ratio of weighed-EDCS as a matching sparsifier compared to [Bernstein-Stein], as well as simple descriptions of a fractional matching and fractional vertex cover defined on top of the EDCS. Considering the wide range of applications EDCS has found in settings such as streaming, sub-linear, stochastic and more we hope our techniques will be of independent research interest outside of the dynamic setting.

\end{abstract}

\section{Introduction.}\label{sec:intro}

Matchings are fundamental objects of combinatorical optimization with a wide range of practical applications. The first polynomial time algorithm for finding a maximum matching was published by Kuhn\footnote{but was attributed to Kőnig and Egerváry}  \cite{Hungarian50} in 1955 which ran in $O(n^4)$ time. A long line of papers have focused on improving this polynomial complexity. Notably Edmonds and Karp \cite{EdmondsK72} showed the first $O(n^3)$ time algorithm for the problem which was later improved to $O(n^{2.5})$ \cite{HopcroftK73}. Mucha and Sankowski \cite{MuchaS04Matrix} showed maximum matching can be solved in matrix multiplication time, that is in $O(n^{\omega})$ where $\omega$ is currently around $2.38$. In the very recent breakthrough result of Chen-Kyng-Liu-Peng-ProbstGutenberg-Sachdeva \cite{ChenKLPGS22}, they showed an $O(m^{1+o(1)})$ time algorithm for the maximum flow problem (which generalizes bipartite matching) providing the first near-linear time algorithm, essentially settling the problem in the sequential setting.

\subparagraph{Dynamic Model.}
This paper focuses on the matching problem in the dynamic model where it has received extensive research attention in recent years, see e.g \cite{BehnezhadK22,Wajc20,ArarCCSW18,BhattacharyaK21,BernsteinS16,PelegS16,AbboudA0PS19,BernsteinFH19,CharikarS18,NeimanS16,Sankowski07,BhattacharyaHN16} and many more. In this setting our task is to maintain a large matching as the graph undergoes updates. We will refer to updates being fully dynamic if they concern both insertions and deletions and partially dynamic if only one of the two is allowed. The objective is to minimize the time spent maintaining the output after each update. Throughout the paper we will always refer to update time in the amortized sense---averaged over the sequence of updates. In \cite{Sankowski07} Sankowski has shown the fist improvement for the fully dynamic maximum matching problem in terms of update time ($O(n^{1.45})$) compared to static recomputation after each update. Unfortunately, based on well accepted hardness conjectures no dynamic algorithm for the maximum matching problem may achieve update time sub-linear in $n$ \cite{HenzingerKNS15Hardness}. Most works focused on the relaxed approximate version of the problem where the goal is to maintain a large matching in $G$ with respect to the maximum matching size $\mu(G)$. We will refer to matching algorithms as $\alpha$-approximate (respectively $(\alpha, \beta)$-approximate) if it maintains matching $M$ such that at all times $|M| \geq \mu(G) \cdot \alpha$ (respectively $|M| \geq \mu(G)  \cdot \alpha - \beta$).

\subparagraph{Fully Dynamic Approximate Matching.} The holy grail of dynamic algorithm design is to achieve an update time of $O(\polylog(n))$ or ideally constant even constant.  For the fully dynamic approximate matching problem, a long line of research \cite{BaswanaGS11, BhattacharyaHN17,BhattacharyaHI15,BhattacharyaK19,BehnezhadDHSS19,OnakR10,Solomon16,BhattacharyaCH17} has lead to algorithms achieving $\approx\frac{1}{2}$-approximation with $O(\polylog(n))$ and constant update time. No fully dynamic algorithm has been found achieving better than $\frac{1}{2}$-approximation in $O(\polylog(n))$ update time for the problem, and this challenge appears to be one of the most celebrated problem within the dynamic matching literature. A set of interesting papers focused on $\approx\frac{2}{3}$-approximation in $\tilde{O}(\sqrt{n})$ update time \cite{BernsteinS15, BernsteinS16, Kiss22ITCS,0001SSU22} and other approximation-ratio to polynomial-update-time trade-offs in the better-than-$\frac{1}{2}$-approximation regime were achieved by \cite{BehnezhadK22,BehnezhadLM20,RoghaniSW22}. Note that through periodic recomputation of the matching (roughly every $\eps\mu(G)$ updates) we can achieve fully dynamic $(1-\epsilon)$-approximation in $\tilde{O}(n)$ update time \cite{GuptaP13}. Very recently \cite{SoheilNEW} has shown that $(1-\eps)$-approximation is possible in slightly sublinear update time $O(n/\log^*(n)^{O(1)})$ suggesting that there might exist efficient non-trivial algorithms for the problem. Note that very recently \cite{behnezhad2023dynamic, bhattacharya2023dynamic} have independently shown that if our goal is to maintain the \emph{size} of the maximum matching (and not the edge-set) then sub-$\frac{1}{2}$ approximation is achievable in polylogarithmic update time.

\subparagraph{Partially Dynamic Matching Algorithms.} For small approximation ratios, achieving polylogarithmic update time for fully dynamic matching seems far out of reach with current techniques, or perhaps even impossible. Hence, a long line of papers have focused on maintaining a ($1-\eps$)-approximate matching in partially dynamic graphs: either incremental (edge insertions) or decremental (edge deletions). The first $O(\poly(\log(n), \eps^{-1}))$ algorithm for maintaining a $(1-\eps)$-approximate matching under edge insertions is due to Gupta \cite{Gupta14Incremental}, with amortized update time $O(\log^2(n)/\eps^4)$. Their algorithm models the bipartite matching problem as a linear program, and uses the celebrated multiplicative-weights-updates framework. Generalizing the result of \cite{Gupta14Incremental} recently Bhattacharya-Kiss-Saranurak \cite{Sayan23DynamicLP} has shown that an arbitrarily close approximation to the solution of a linear program undergoing updates either relaxing or restricting (but not both) its solution polytope can be maintained in $O(\poly(\log n, \eps^{-1}))$ update time. Hence, the algorithm of \cite{Sayan23DynamicLP} shows how to maintain a $(1-\eps)$-approximate matching under either decremental or incremental updates with a unified approach. As both of these papers rely on static linear program solver sub-routines their update times inherently carry $\log(n)$ factors, and it seems implausible that these techniques can achieve constant update time independent of $n$.

The decremental algorithms of \cite{BernsteinGS20,JambulapatiJST22} focus on maintaining ``evenly spread out'' fractional matchings so that they are robust against edge-deletions. These algorithms rely on either maximum-flow computation or convex optimization sub-routines which similarly to LP-solvers carry $\log(n)$-factors into the update time.

The first constant time\footnote{That is constant-time whenever $\eps$ is constant, i.e. the update time should be independent of $n$.} partially dynamic matching algorithm is due to \cite{Grandoni19Exponential} who solve $(1-\eps)$-approximate matching in incremental graphs with an update time of $(1/\eps)^{O(1/\eps)}$. Their solution relies on augmenting path elimination, a technique used commonly for the matching problem in the static setting. However, enumerating augmenting paths seems to inherently carry an exponential dependency on $1/\eps$ due to the number of possible such paths present in the graph.

Very recently Zheng-Henzinger \cite{Zheng23IPCODecrementalNew} has shown the first $O(\poly(1/\eps))$ update time algorithm for maintaining a $(1-\eps)$-approximate matching in decremental bipartite graphs. They note that the celebrated and rather simple auction based matching algorithm \cite{auction1, auction2} almost seamlessly extends to the setting of edge deletions. However, for incremental
bipartite graphs the previous techniques either require update time that is not constant (dependency on $\log n$) or with exponential dependency on $1/\eps$ (unlike for decremental, where $O(\poly(1/\eps))$ update time is achievable).
These developments leave the following apparent gap in the literature of partially dynamic matching algorithms:

\begin{wrapper}

\label{q:1}

\textbf{Question 1}: Can we maintain a $(1-\eps)$-approximate maximum matching of a bipartite graph as it undergoes partially dynamic edge updates in $O(\poly(\eps^{-1}))$ update time?

\end{wrapper}

Based on the current landscape of the dynamic algorithms literature, achieving $(1 - \eps)$-approximation under fully dynamic updates in small update times seems to be far out of reach. Contrary to the extensive research effort, no fully dynamic algorithm with $\poly(\log(n), \eps^{-1})$ has been found for maintaining matchings with better than $\frac{1}{2}$-approximation. This apparent difficulty proposes the research of dynamic models somewhere between fully and partially dynamic updates. The previously mentioned paper by Zheng-Henzinger \cite{Zheng23IPCODecrementalNew} implements a $(1-\epsilon)$-approximate algorithm with $O(1/\eps)$-update time which can support vertex insertions and deletions on separate sides of the bipartition. The existence of this new result proposes the following natural question:

\begin{wrapper}

\textbf{Question 2:} Under what kind of non-partially dynamic updates can we maintain a $(1-\eps)$-approximate maximum matching of a bipartite graph?
    
\end{wrapper}

\subsection{Our Contribution}

In this paper we provide positive answers to both \textbf{Question 1} and \textbf{Question 2}. Our main result is the first $O(\poly(1/\eps))$ update time $(1-\eps)$-approximate dynamic matching algorithm for bipartite graphs undergoing edge insertions:

\begin{restatable}{theorem}{mainedgeupdates}

\label{thm:integral-matching-algo}

There exists a deterministic dynamic algorithm which for arbitrary small constant $\eps > 0$ maintains a $(1-\eps)$-approximate maximum matching of a bipartite graph undergoing edge insertions in total update time $O(n/\eps^{6} + m/\eps^{5})$.
\end{restatable}

Previous algorithms for $(1-\eps)$ approximate dynamic matching under edge updates required update times which were either super-constant \cite{Gupta14Incremental,Sayan23DynamicLP} or had an exponential dependency on $\eps^{-1}$ \cite{Grandoni19Exponential}. Furthermore, our algorithm is arguably simpler then previous implementations and it is self contained (except for the static computation of ($1-\eps$)-approximate maximum matchings) where as most dynamic matching algorithms either rely on heavy machinery from previous papers or use black-box tools like multiplicative weight updates or flow-subroutines.

We further show that if we allow for (additive) $(1, \eps \cdot n)$-approximation\footnote{Recall that this means that we maintain a matching of size at least $\mu(G)-\eps n$, as opposed to $\mu(G)-\eps\mu(G)$} our algorithm seamlessly extends to a wider range of updates:

\begin{restatable}{theorem}{mainvertexupdates}
\label{thm:vertex-deletions-integral-matching-algo}

There exists a deterministic dynamic algorithm which for arbitrary small constant $\eps > 0$ maintains a $(1, \epsilon \cdot n)$-approximate maximum matching of a bipartite graph undergoing edge insertions and vertex deletions in total update time of $O(n/\eps^{8} + m/\eps^{5})$.

\end{restatable}

In contrast to the similar update time result of \cite{Zheng23IPCODecrementalNew} which allows for edge deletions and vertex insertions on one side of the bipartition our algorithm allows from arbitrary vertex deletions. Our algorithm maintains a $(1-\eps)$-approximate maximum matching of the graph throughout updates which can both increase and decrease the maximum matching size of the graph. Hence, we hope our techniques provide useful insight towards fully dynamizing $(1-\eps)$-approximate algorithms for the matching problem.

Our algorithm relies on the weighted variant of the celebrated Edge-Degree-Constrained-Subgraph (EDCS) matching sparsifier. The unweighted EDCS (first introduced by Bernstein-Stein \cite{BernsteinS15}) has found applications in a number of different computational settings: streaming \cite{Bernstein20RandomStream, RandomLarge, BeatingRandom}, stochastic, one way communication, fault tolerant \cite{TowardsUNified}, sub-linear \cite{behnezhad2023dynamic, bhattacharya2023dynamic, sublinearSoheil,sublinearSayan} and dynamic \cite{BernsteinS15, BernsteinS16, Kiss22ITCS,0001SSU22,behnezhad2023dynamic}. On the other hand the \emph{weighted} EDCS variant which provides a tighter approximation has only found applications in small arboricity graphs \cite{BernsteinS15}. Hence, we initiate the study of the weighted EDCS in dense graphs. 

Furthermore, we show a significantly simplified proof for the approximation ratio of the weighted EDCS datastructure with respect to the maximum matching size. In our proof, we identify simple and explicit descriptions of a fractional matching and fractional vertex covers defined on top of the weighted EDCS, which might be of independent interest. Moreover, we show that the dependence on the slack parameter on the maximum degree of the weighted EDCS is exactly quadratic. This in sharp contrast to the unweighted EDCS where the same relationship have been proven to be linear \cite{SOheilTIghtAnalysis}. While within the dynamic matching algorithm literature papers don't tend to focus on exact $\epsilon$ complexities but rather $n$ dependence, in models such as semi-streaming and distributed the $\epsilon$ dependency usually gains more focus. Our hard example (most likely) rules out applications of weighted EDCS in these models for obtaining sub $\epsilon^{-2}$ round/pass complexity algorithms. We hope that these observations will be of independent research interest due to the wide-spread popularity of the EDCS datastructure for solving matching problems.

\subsection{Our Techniques}

Assume $H$ is a multi-graph defined on the vertex set of $G$ and let $\dH(v)$ stand for the degree of vertex $v$ in $H$. We define the degree of an edge to be the sum of the degrees of its endpoints. \footnote{Note that we are sticking to the notation weighted-EDCS instead of multi-EDCS to be in line with the naming convention of \cite{BernsteinS15} which defined $H$ to be a weighted graph with integer edge weights.}

\begin{wrapper}

\begin{definition}[{Weighted EDCS \cite{BernsteinS15}}]
\label{def:wedcs}
 Given a graph $G = (V,E)$, a multiset $H$ is called a weighted EDCS with parameter $\beta$ if\footnote{Some authors use an additional parameter $\beta^{-} < \beta$ which replaces the ``$\beta-1$'' in the degree-constraint. For our purposes, we will always have $\beta^{-}=\beta-1$}:
\begin{enumerate}
    \item $\dH(u)+\dH(v) \le \beta$ for all edges $(u,v)\in H$.
    \item $\dH(u)+\dH(v) \ge \beta-1$ for all edges $(u,v)\in E$.
\end{enumerate}
\end{definition}
If $H$ is not a weighted EDCS, we call an edge $e\in H$ \emph{overfull} if it violates (i), and an edge $e\in E$ \emph{underfull} if it violates (ii).

\end{wrapper}

If $\beta = \Omega(\eps^{-2})$ and $H$ is a $\beta$-WEDCS of $G$ then $\mu(H) \geq \mu(G) \cdot (1-\eps)$ (\cite{BernsteinS15}, Theorem~\ref{thm:explicit-fractional-matching}). In order to derive our incremental result we show that a $\beta$-WEDCS can be efficiently maintained greedily under edge insertions. In turn we can efficiently maintain a $(1+\epsilon)$-approximate matching within the support of $H$ through periodic recomputation. 

Define a valid update of $H$ to be one of the following: (i) and edge $e \in H$ which is overfull with respect to $H$ gets deleted from $H$ (ii) a copy of an edge $e \in E$ which is underfull with respect to $H$ is added to $H$. In \cref{clm:updates} (slightly improving on the similar lemma of \cite{BernsteinS15}) we show that if $H$ is initialized as the empty graph and may only undergo valid updates it may undergo a total of $O(\mu(G) \cdot \beta^2)$ updates. 

Fix some $\beta = \Theta(\eps^{-2})$. Assume $G$ is initially empty initialize $H$ to be an empty edge set (note that by definition initially $H$ is a $\beta$-WEDCS of $G$). Assume edge $e$ is inserted into $G$. If at this point $e$ is not underfull with respect to $H$ there is nothing to be done as $H$ remained a valid WEDCS of $G$. If $e$ is underfull with respect to $H$ we add copies of it to  $H$ until it is not. This process of adding $e$ to $H$ has increased the edge degree of edges neighbouring $e$ in $H$ and some of them might have became overfull. To counteract this we iterate through the neighbours of $e$ in an arbitrary order and if we find an overfull edge $e'$ we remove it from $H$. This edge removal decreases the edge degrees in the neighbourhood of $e'$. To counteract this we recurse and look for underfull edges in the neighbourhood of $e'$. If such an edge $e''$ is found we add copies of it to $H$ until $e''$ is not underfull and repeat the same steps as if $e''$ was just inserted into $G$. This defines a natural recursive process for restoring the WEDCS properties after each edge insertion in a local and greedy way.

Whenever we have to explore the neighbourhood of an edge in $O(\Delta)$ time (where $\Delta$ is the max-degree) to either check for underfull or overfull edges we do so because $H$ underwent a valid update. By \cref{clm:updates} this may only happen at most $O(\mu(G) \cdot \beta^2)$ times. Hence, naively the total work spent greedily fixing the WEDCS properties is $O(\mu(G) \cdot \Delta \cdot \beta^2)$. For some graphs this value might be significantly larger then $m$. In order to improve the update time to $O_\beta(m)$ we assign a counter $c_v$ to each vertex $v$ measuring the number of valid updates of $H$ the neighbourhood of $v$ has underwent. Once $c_v$ grows to $\Omega(\beta^2 \cdot \eps^{-1})$ we mark $v$ dirty and ignore further edges inserted in the neighbourhood of $v$. By marking a single vertex dirty and ignoring some edges incident on it we may loose out only on a single edge of any maximum matching. However, whenever we mark a vertex dirty we can charge $\Omega(\beta^2 \cdot \eps^{-1})$ valid updates of $H$ to that vertex. As there may be at most $O(\mu(G) \cdot \beta^2)$ valid updates of $H$ in total we may only mark $O(\mu(G) \cdot \eps)$ vertices dirty hence we will only ignore an $O(\epsilon)$-fraction of any maximum matching within the graph through ignoring edges incident on dirty vertices. As we may scan the neighbourhood of vertex $v$ at most $O(\beta^2 \cdot \eps^{-1}) = \eps^{-O(1)}$ times until $v$ is marked dirty we ensure that each edge is explored $\eps^{-O(1)}$ times guaranteeing a total running time of $O(m \cdot \eps^{-O(1)})$. Full details can be found in \cref{sec:algo}.

\subparagraph{Towards Full Dynamization.} The algorithm almost seamlessly adopts to vertex deletions if we allow for $(1, \eps \cdot n)$-approximation\footnote{Readers may reasonable argue that the additive slack is not necessary as a number of vertex-sparsification techniques exist in literature allowing us to improve the approximation to purely multiplicative slack in the dynamic setting. Unfortunately, these techniques don't appear to be robust against vertex-wise updates}. Whenever a vertex gets deleted from the graph our WEDCS $H$ might be locally affected. This means that over the full run of the algorithm, $H$ may undergo further valid updates then the $O(\mu(G) \cdot \beta^2)$ bound provided by \cref{clm:updates}. A potential function based argument allows us to claim that each vertex deletion may increase the total number valid updates $H$ may undergo by $O(\beta^2)$. As each vertex may be deleted at most once this means that the total number of valid updates we might make to restore $H$ is $O(n \cdot \beta^2)$, each update requiring $O(\Delta)$ time if naively implemented. By marking vertices as dirty as before we can guarantee amortized $O(\poly(1/\eps))$ update time. However, now we must mark up to $\approx \eps\cdot n$ vertices as dirty (as opposed to $\approx \eps \cdot  \mu(G)$ like before), which means we may miss out on $\eps \cdot n$ edges of the maximum matching.

\subsection{Acknowledegemnts}

The authors would like to acknowledge Danupon Nanongkai for helpful discussions.


\section{Preliminaries.}\label{sec:prelim}

\subparagraph{Matching Notation.} Let $N_E(v)$ stand for the edges neighbouring vertex $v$ in $E$. A fractional matching $f$ of a graph $G$ is an assignment of the edges of $G$ to values in the range $[0,1]$ such that for all vertices $v \in V$ it holds that $\sum_{e \in N_E(v)} f_e \leq 1$. The \emph{size} of a fractional matching is simply the sum of the fractional values over its edges. That is a maximum fractional matching is the solution to the linear program $\max \{\sum_{e\in E} f_e : \sum_{e\in N_E(v)} f_e \le 1 \text{ for all $v\in V$}, f\ge 0\}$.
A solution $x$ to the dual of this program
$\min\{\sum_{v\in V} x_v : x_u+x_v \ge 1 \text{ for all $(u,v)\in E$}, x\ge 0\}$ 
is a fractional vertex cover.

Approximation with respect to a fractional matching is defined simmilarly as with respect to integral matchings. For a graph $G = (V,E)$ we use $\mu(G)$ to denote the size of the maximum matching in $G$. Likewise, we use $\mu^*(G)$ to denote the size of the maximum \emph{fractional} matching. It is well-known that $\mu(G)\le \mu^*(G)\le \frac{3}{2}\mu(G)$ for any graph, and that $\mu^*(G) = \mu(G)$ in bipartite graphs. 

\begin{theorem}[{Hopcroft-Karp \cite{HopcroftK73}}]
\label{thm:staticapproxmatching}
There exists a deterministic static algorithm which finds a $(1-\eps)$-approximate maximum matching of a graph $G$ on $m$ edges in $O(m / \eps)$ time. 
\end{theorem}

\section{Incremental Approximate Matching.}  \label{sec:algo}
We start by showing our incremental fractional matching algorithm, and then show how to extend it (for bipartite graphs) to also maintain an integral matching.

\subsection{Weighted EDCS \& Fractional Matchings.} 
\label{sec:algo-frac}

In this section we show our algorithm to (almost\footnote{As we will see later in this section, our sparsifier $H$ will be a weighted EDCS for $G\setminus R$, where $R$ is a subgraph of $G$ with very small maximum matching size $\mu(R) = O(\eps \mu(G))$.}) maintain a weighted EDCS $H$ in an incremental graph.
It is well-known that such an $H$ will be a $(1-\eps)$-matching sparsifier on \emph{bipartite graphs}, that is a ``sparse'' subgraph with
$\mu(H) \ge (1-\eps)\mu(G)$ \cite{BernsteinS15}.

As we show later in \cref{sec:explicit-fractional-matching}, we even known something stronger: there is an explicit fractional matching in $H$ of size at least $(1-\eps)\mu^*(G)$, defined as
\begin{equation}
\label{eq:frac}
f_{(u,v)} = \min\left(\frac{1}{\dH(v)},\frac{1}{\dH(u)}\right) \text{ on each } (u,v)\in H.
\end{equation}
This fractional matching is also valid for general (non-bipartite) graphs. Hence our incremental algorithm will also maintain this explicit $(1-\eps)$-approximate fractional matching (even in non-bipartite graphs). Formally we prove the following theorem.

\begin{theorem}

\label{thm:fractional-matching-algo}
For any $\eps\in (0,1)$, there is an algorithm (\cref{alg:incremental-wedcs}) that maintains a $(1-\eps)$-approximate maximum \emph{fractional} matching in an incremental graph in total update time $O(n/\eps^6 + m/\eps^5)$. Additionally, this fractional matching is always supported on a set of edges $H$ of size $|H|\le \Theta(\mu(G)/\eps^2)$ and maximum degree $O(1/\eps^2)$.

\end{theorem}

First we need two standard facts about weighted EDCS. For completeness, we prove these in \cref{appendix:proofs}. \cref{clm:updates} has only been shown before for unweighted EDCS \cite{BernsteinS15} and not weighted (but the arguments are very similar).

\begin{restatable}{lemma}{clmdegrees} \label{clm:degrees}
In a $\beta$-WEDCS $H$, the maximum degree is at most $\beta$ and $|H|\le \beta\mu^*(G)$.
\end{restatable}

\begin{restatable}{lemma}{clmupdates} \label{clm:updates}
If a multiset of edges $H$ is only ever changed by removing overfull edges and adding underfull edges, then there are at most $\beta^2\mu^*(G)$ such insertions/deletions to $H$.
\end{restatable}

\subparagraph{Overview.}
Our algorithm (see \cref{alg:incremental-wedcs}) will maintain a weighted EDCS $H$ with $\beta = \Theta(1/\eps^2)$.
We also maintain the $(1-\eps)$-approximate fractional matching $f$ as in \cref{eq:frac} and \cref{thm:explicit-fractional-matching}. 

When we get an edge-insertions $(u,v)$, we need to reestablish the property that $H$ is an EDCS. If $(u,v)$ is underfull $(\dH(u)+\dH(v)<\beta-1)$, we add it (maybe multiple times) to $H$. This means that $\dH(u)$ (similarly $\dH(v)$) increases, which can potentially make some incident edge $(u,w)\in H$ overfull $(\dH(u)+\dH(w)>\beta)$, so we must remove one such edge. This might in turn
lead to some edge $(w,z)\in E$ being underfull (as now $\dH(w)$ decreased), so we add this edge to $H$. This process continues, so both from $u$ and $v$ we need to search for alternating paths of underfull and overfull edges (as is standard in EDCS-based algorithms). In total, \cref{clm:updates} says there are $O(n/\eps^4)$ updates to $H$ over the full run of the algorithm.

We note that searching for an \emph{overfull} edge is cheap: the maximum degree in $H$ is just $O(\beta)$ (\cref{clm:degrees}), so we can afford to, in $\Theta(1/\eps^2)$ time, check all incident edges. However, searching for \emph{underfull} edges is more expensive: this time we cannot afford to just go through all neighboring edges in $E$, as we no longer have a bound on the maximum degree.

To overcome this we use an amortization trick which allows us to ignore a vertex if we touched it too many times. There are only $\beta^2\mu^*(G)$ updates to $H$ in total (\cref{clm:updates}), so there will only be $\eps \mu^*(G)$ many vertices incident to more than $2\beta^2/\eps$ of these updates. Any edges incident to these ``update-heavy'' vertices we may ignore, as this may only decrease the maximum matching size by an $\eps$ fraction. We thus only need to check each edge a total of $O(1/\eps^5)$ times over the run of the algorithm, except when it is in $H$ already.
Note that $H$ is no longer a weighted EDCS of $G = (V,E)$, but rather of $G'= (V,(E\setminus R)\cup H)$ where $R$ is this set of edges we ignored (with $\mu^*(R)\le \eps \mu^*(G)$).

\SetKwFunction{FixEdge}{FixEdge}
\SetKwFunction{FixVertex}{FixVertex}
\SetKwData{Visits}{visits}
\begin{algorithm}[!ht]
\caption{Incremental Weighted EDCS \& Frasctional Matching}\label{alg:incremental-wedcs}
\DontPrintSemicolon
\SetKwProg{Fn}{function}{}{}

\tcp{Initially $E = H = \emptyset$ and $\dH(v) = \Visits[v]= 0$ for all $v\in V$.}
\tcp{When an edge insertion $e$ appears, add it to $E$ and call \FixEdge{$e$}.}
\BlankLine

\Fn{\FixEdge{$e=(u,v)$}}{
    \If(\tcp*[f]{overfull}){$\dH(u)+\dH(v) > \beta$ and $(u,v)\in H$}{
        Remove (one copy of) the edge $(u,v)$ from $H$\;
    }
    \If(\tcp*[f]{underfull}){$\dH(u)+\dH(v) < \beta-1$}{
        Add (one copy of) the edge $(u,v)$ to $H$\;
    }
    \If{the edge was added or removed}{
        Update $\dH(u), \dH(v)$, and the fractional matching accordingly\;
        \FixVertex{$u$}, \FixVertex{$v$}\;
    }
}
\BlankLine
\Fn{\FixVertex{$v$}}{
    $\Visits[v] \gets \Visits[v]+1$\;
    \If{$\Visits[v] < 2\beta^2/\eps$}{
        \For{edge $e\in E$ incident to $v$}{
            \FixEdge{$e$}\; \label{alg:line:all-edges}
        }
    }
    \Else{
        \For{edge $e\in H$ incident to $v$}{
            \FixEdge{$e$}\;
        }
    }
}
\end{algorithm}


\subparagraph{Running Time.} We analyse the total update time spent in different parts of our algorithm. 
\begin{itemize}
\item We first note that \FixEdge runs in constant time whenever it does not update $H$. 
It is called once for each edge-insertion (total $O(m)$ times), and also some number of times from \FixVertex.

\item Now consider the case when \FixEdge does update $H$ (which happens at most $\beta^2\mu^*(G)$ times per \cref{clm:updates}). Now the algorithm uses $O(\beta)$ time for the update of the fractional matching and insertion/removal in $H$, in addition to exactly two calls to \FixVertex. Except for these calls to \FixVertex, over the run of the algorithm we hence spend a total of $O(\mu(G)\beta^3) = O(n/\eps^6)$ time.

\item By the previous point, we will call \FixVertex at most $2\beta^2\mu^*(G)$ times. In each call, we either loop through all incident edges in $H$ or $E$. If we loop through $H$, we visit $\beta$ many edges (by \cref{clm:degrees}). Either these \FixEdge calls take constant time, or they are already accounted for in the previous point. In total, this thus accounts for another $O(\mu(G)\beta^3) = O(n/\eps^6)$ running time.

\item We account for the case when \FixVertex loops through all incident edges in $E$ differently. Consider how often a specific edge $e$ appears in the for-loop at line~\ref{alg:line:all-edges}. Each endpoint vertex of $e$ will reach this line at most $O(\beta^2/\eps)$ times. Hence, in total for all edges, line~\ref{alg:line:all-edges} is run at most $O(m\beta^2/\eps) = O(m/\eps^5)$ times.
\end{itemize}

\subparagraph{Approximation Guarantee.} We now argue the approximation ratio. We will show that the fractional matching supported on $H$ is a $(1-2\eps)$-approximation of maximum fractional matching in $G$. If one want a $(1-\eps')$-approximation, then one can run the algorithm in the same asymptotic update time setting $\eps = \eps'/2$, and changing $\beta$ accordingly.

Define $R_V$ to be the set of ``update-heavy'' vertices: that is vertices $v$ for which $\FixVertex{v}$ has been called at least $2\beta^2/\eps$ many times (i.e.\ $\Visits[v] \ge 2\beta^2/\eps$). By a counting argument $|R_V|\le 2\beta^2\mu^*(G) / (2\beta^2/\eps) = \eps\mu^*(G)$, since by \cref{clm:updates} in total there are only $\beta^2\mu^*(G)$ many updates to $H$, each issuing exactly two calls to $\FixVertex$.
If $R_E$ is the set of edges incident to $R_V$, then $\mu^*(R_E)\le |R_V|\le \eps\mu^*(G)$ since $R_V$ is a vertex cover of $R_E$. 

Define $G' = (V,E\setminus (R_E\setminus H))$. By the above, $\mu^*(G')\ge (1-\eps)\mu^*(G)$. We will finish the proof by arguing that $H$ is a weighted EDCS of $G'$, and thus that our fractional matching is of value at least $(1-\eps)\mu^*(G') \ge (1-\eps)^2 \mu^*(G) \ge (1-2\eps)\mu^*(G)$. Whenever an edge is added or removed to $H$, we call $\FixVertex$ on its endpoints, and no other degrees $\dH$ have changes. Every time $\FixVertex(v)$ is called for $v\not\in R_V$, we make sure that all edges $e\in E$ incident to it satisfy the definition of an EDCS, and when $\FixVertex{v}$ is called for some $v\in R_v$, we check the edges incident to $H$. We note that when a vertex becomes ``update-heavy'' (added to $R_v$),
then we do \textbf{not} immediately remove all incident edges from $H$ (as then we no longer have the same bound on the number of updates to $H$ since \cref{clm:updates} no longer applies).

\begin{remark}
We note that our algorithm runs in time $O(n\beta^3 + m\beta^2/\eps)$, and a valid question is whether setting $\beta = \Theta(1/\eps^2)$ is actually necessary? Recently it was shown that for unweighted EDCS $\beta = \Theta(1/\eps)$ is enough \cite{SOheilTIghtAnalysis}. However, for weighted EDCS the $\eps^2$ dependency is indeed necessary, as we show by an example where this is asymptotically tight in \cref{sec:lower-bound}.
\end{remark}

\subsection{Integral Matchings in Bipartite Graphs.} 
\label{sec:algo-int}
In this section we argue how to extend our fractional matching algorithm (\cref{thm:fractional-matching-algo}) to maintain an integral matching instead (for bipartite graphs), in the same asymptotic update time. We cannot use known dynamic rounding techniques \cite{Sayan23DynamicLP, Wajc20}, since all these incur $\polylog(n)$ factors or require randomization, and we are aiming for update time independent of $n$. In fact, our technique is simple and combinatorial; and only relies on the standard Hopcroft-Karp algorithm for finding a $(1-\eps)$-approximate matching in the static setting \cite{HopcroftK73}.

\mainedgeupdates*

\begin{remark}
\label{remark:gupta-peng-framework}
Before proving the above theorem, we briefly explain how to achieve a slightly less efficient version (amortized $O(1/\eps^8)$ update time) by using the fully dynamic $(1-\eps)$-approximate matching algorithm of Gupta-Peng \cite{GuptaP13} as a black box.
The idea is to run the fully dynamic algorithm on our sparsifier---the weighted EDCS $H$. This way we maintain a matching $M$ of size $|M| \ge (1-\eps) \mu(H) \ge (1-\eps)^2 \mu(G) \ge (1-2\eps) \mu(G)$.

Gupta-Peng \cite{GuptaP13} state that their algorithm runs in $O(\sqrt{m}/\eps^2)$ time per update. However, as previously pointed out by e.g. \cite[Lemma~1]{BernsteinS15}, it is in fact more efficient when the max-degree $\Delta$ is low, in which case the update time is only $O(\Delta/\eps^2)$. Since $H$ always has max-degree $\beta = \Theta(1/\eps^2)$, we can maintain the integral matching $M$ in $O(1/\eps^4)$ time \emph{per update to $H$}. Over the run of the algorithm, we only perform $O(\mu(G)/\eps^{4})$ updates to $H$ (see \cref{clm:updates}), hence the total additional update time spent maintaining the integral matching will be $O(n/\eps^{8})$.
\end{remark}

\begin{proof}[Proof of \cref{thm:integral-matching-algo}.]
To prove \cref{thm:integral-matching-algo}, we need a slightly more refined analysis than the one above. We still run our incremental algorithm (\cref{alg:incremental-wedcs,thm:fractional-matching-algo}) to maintain a weighted EDCS $H$ together with a fractional matching supported on $H$. Similarly to above, we additionally maintain an $(1-\eps)$-approximate (integral) matching $M$ of $H$.

The main idea of the fully dynamic algorithm of Gupta-Peng \cite{GuptaP13} is to lazily recompute (in $O(|H|/\eps)$ time via Hoproft-Karp Theorem~\ref{thm:staticapproxmatching}) $M$ every $\approx \eps\mu$ updates to $H$ (indeed, during this few updates, the matching size cannot change its value by more than $\eps\mu$). There are two observations which helps us to do better:
\begin{enumerate}
\item The graph $G$ (but not the sparsifier $H$) is incremental, so $\mu(G)$ can only grow.
\item We know a good estimate of $\mu(G)$, namely the size of our fractional matching. Denote by $\tilde{\mu}$ the value of the maintained fractional matching, so that $(1-\eps)\mu(G) \le \tilde{\mu} \le \mu(G)$.
\end{enumerate}
The above two observations mean that we only need to recompute the matching $M$ whenever $\mu(G)$ actually have increased significantly (namely by a $(1+\Theta(\eps))$-factor), and not just every $\eps \mu$ updates.

Formally, whenever $|M| \ge (1-\eps)^2 \tilde{\mu}$ we know that $M$ is still a $(1-3\eps)$-approximation since then $|M|\ge (1-\eps)^2\tilde{\mu} \ge (1-\eps)^3\mu(G) \ge (1-3\eps)\mu(G)$.
Conversely, whenever $|M| < (1-\eps)^2 \tilde{\mu}$, we recompute $M$ in time $O(|H|/\eps)$ (Theorem~\ref{thm:staticapproxmatching}) so that it is a $(1-\eps)$-approximation of the maximum matching in $H$ (and thus also a $(1-2\eps)$-approximation of the maximum matching in $G$).

Let us now bound the total time spent recomputing $M$. Let $M_1, M_2, \ldots, M_t$ be the different approximate matchings we compute during the run of the algorithm.
We first note that at the time when we compute $M_{i+1}$:
\begin{equation}
|M_i| \le (1-\eps)^{2} \tilde{\mu} \le 
(1-\eps)\left((1-\eps) \mu(H)\right) \le (1-\eps) |M_{i+1}|
\end{equation}
This in turn means that $|M_{i}| \le (1-\eps)^{t-i}\, n$ (since $|M_t| \le n$), and hence that
$\sum_{i=1}^{t} |M_i| \le n \sum_{i=0}^\infty (1-\eps)^i \le n/\eps$, by a geometric sum.

Finally we note that we spend $O(|M_i|/\eps^3)$ time in order to compute $M_i$. Indeed,
when we compute $M_i$, we did so in $O(|H|/\eps)$ time, and $|H| = O(\mu(G)/\eps^2)$ by \cref{clm:degrees}.
This means that in total, over the run of the algorithm, we spend $O(\sum |M_i|/\eps^3) = O(n/\eps^4)$ time maintaining the integral matchings $M_i$. This is in addition to the time spent maintaining the weighted EDCS $H$ and the fractional matching (see \cref{thm:fractional-matching-algo}). This concludes the proof of \cref{thm:integral-matching-algo}.
\end{proof}

\section{Vertex Deletions.}
\label{sec:vertex-deletions}
In this section we will observe that our algorithm can also handle vertex deletions (simultaneously to handling edge insertions) in similar total update time. However, this comes with one caveat: we instead get additive approximation error proportional to $\eps n$ (that is we maintain a matching of size $\mu^*(G)-\eps n$, instead of $\mu^*(G)-\eps \mu^*(G)$ as before). 


\begin{theorem}
\label{thm:vertex-deletions-fractional-matching-algo}
For any $\eps\in (0,1)$, there is an algorithm  that maintains a \emph{fractional} matching of size at least $\mu^*(G)-\eps n$ in an graph $G$ undergoing edge insertions and vertex deletions. The total update time is $O(n/\eps^6 + m/\eps^5)$.
\end{theorem}
\begin{proof}
The algorithm (\cref{alg:incremental-wedcs}) remains the same as in \cref{sec:algo-frac}. When a vertex is deleted, we simply remove all it's incident edges from $E$ and $H$, and call $\FixVertex$ on all neighboring vertices (in $H$) who now changed their degree. The only thing which changes in the analysis is the $\beta^2\mu^*(G)$-bound on the number of updates to $H$ (\cref{clm:updates}), which no longer applies. However, we can still get a weaker version of \cref{clm:updates} with a $\beta^2 n$ total update bound instead:
\begin{restatable}{claim}{clmupdatesvertex} \label{clm:updates-vertex}
If a multiset of edges $H$ is only ever changed by (i) removing overfull edges, (ii) adding underfull edges, and (iii) \emph{removing all edges incident to a vertex when no edges are underfull or overfull}, then there are at most $3\beta^2 n$ insertions/deletions to $H$.
\end{restatable}

Given \cref{clm:updates-vertex} (which we prove in \cref{appendix:proofs}), we see that the running time analysis of \cref{alg:incremental-wedcs} can remain exactly the same! In the approximation guarantee analysis, we now have more ``update-heavy'' vertices $|R_V| \le 6\beta^2 n / (2\beta^2/\eps) = 3\eps n$, which is why we now can lose up to $O(n\eps)$ edges from the matching. Otherwise, the approximation guarantee analysis remains the same, and so does the rest of the analysis of the algorithm.
\end{proof}

\subparagraph{Rounding in Bipartite Graphs.} Similar as in \cref{sec:algo-int}, we can round the fractional matching to an integral one in bipartite graphs, also while supporting edge-insertions and vertex-deletions simultaneously.

\mainvertexupdates*

\begin{proof}
Unlike in \cref{sec:algo-int}, we cannot argue that $\mu(G)$ is increasing when we have vertex deletions. So instead we resort to the Gupta-Peng \cite{GuptaP13} framework discussed in \cref{remark:gupta-peng-framework} (together with \cref{clm:updates-vertex}), which has the additional cost of $O(n/\eps^8)$ total update time to maintain an approximate integral matching on our sparsifier $H$.
\end{proof}

\begin{remark}
We note that normal vertex-sparsification techniques (such as the one shown in \cite{Kiss22ITCS} against oblivious adversaries) do not apply here in order to assume $n = \tilde{\Theta}(\mu(G))$ so that the additive error becomes multiplicative again. This is because vertex deletions in the original graph might become edge deletions in the vertex-sparsified graph.
\end{remark}
%
%

\section{Tight Bounds of the of Approximation Ratio of a Weighted EDCS.}

\subsection{Explicit Fractional Matching}
\label{sec:explicit-fractional-matching}

 In this section, we give explicit formulas only based on the degrees in $H$, for a $(1-\eps)$-approximate fractional matching. We prove this by also providing an explicit approximate fractional vertex cover, and showing that these satisfy approximate complimentary slackness. This also significantly simplifies the previous proof \cite{BernsteinS15} that $H$ is $(1-\eps)$-matching sparsifier in bipartite graphs.

\begin{theorem}
\label{thm:explicit-fractional-matching}
Suppose $H$ is a weighted EDCS of a graph $G$, with parameter $\beta \ge 36/\eps^2$.
Let $f_{(u,v)} = \min\{\frac{1}{\dH(v)},\frac{1}{\dH(u)}\}$ for each\footnote{We note that edges $e$ appearing multiple time in $H$ all contribute towards $f_{e}$: if $e$ appears $\phi_e$ times in $H$, the value of $f_{e}$ is naturally scaled by $\phi_e$.} $(u,v)\in H$. Then $f$ is a $(1-\epsilon)$-approximate fractional matching of $G$.
\end{theorem}
Define $r_v := \dH(v)-\frac{\beta-1}{2}$ for a vertex $v\in H$.
We define the fractional matching $f$ as in the statement of the theorem, together with the fractional vertex cover $x$:
\begin{align}
f_{(u,v)} &= \min \left( \tfrac{1}{\dH(u)}, \tfrac{1}{\dH(v)}\right) \quad \text{for edge $(u,v)\in H$} \\
x_{v} &= \begin{cases}\min(1, \tfrac{1}{2}+\tfrac{r_v^2}{\beta}) & \text{if $r_v\ge 0$} \\ \max(0, \tfrac{1}{2}-\tfrac{r_v^2}{\beta}) & \text{if $r_v < 0$}\end{cases}
\end{align}
It is now relatively straightforward (albeit a bit calculation-heavy) to argue that $f$ and $x$ are indeed feasible solutions and that they satisfy approximate complimentary slackness.
\begin{claim}
Our $f$ is a fractional matching and our $x$ is a fractional vertex of $G$.
\end{claim}
\begin{proof}
Our $f$ is feasible since no vertex $v$ is overloaded by the matching: at most $\dH(v)$ many incident edges to $v$ contribute at most $1/\dH(v)$ each.

To argue that $x$ is feasible, consider some edge $(u,v)\in E$. Without loss of generality we may assume that $\dH(v)\ge \dH(u)$ and $\dH(v) \ge \frac{\beta-1}{2}$, i.e. $r_v\ge r_u$ and $r_v\ge 0$ (since $\dH(u)+\dH(v)\ge \beta-1$ as $H$ is a weighted EDCS). If $r_v^2 \ge \beta$, $x_v = 1$ so $(u,v)$ is covered. In the case $r_v^2 < \beta$, we instead have $x_v = \frac{1}{2}+\frac{r_v^2}{2}$. It is always the case that $x_u \ge \frac{1}{2}-\frac{r_u^2}{2}$. Additionally we note that $r_u + r_v \ge 0$ (so $r_u^2 \le r_v^2$) since $\dH(u)+\dH(v)\ge \beta-1$, so we conclude that
$x_u + x_v \ge 1$, and hence that $(u,v)$ is covered.
\end{proof}

\begin{claim}
The fractional matching $f$ and fractional vertex cover $x$ satisfy\\ $(1-\frac{4}{\sqrt{\beta}}, 1+\frac{2}{\sqrt{\beta}}+\frac{1}{\beta})$-approximate complementary slackness\footnote{For completeness, we define the approximate complimentary slackness conditions in \cref{appendix:proofs} and prove them in \cref{lem:approx-cs}.}; in particular:
\begin{enumerate}
    \item Whenever $f_{(u,v)} > 0$, then $x_u+x_v \le 1+\frac{2}{\sqrt{\beta}}+\frac{1}{\beta}$.
    \item Whenever $x_{v} > 0$, then $\sum_{u : (u,v)\in H}f_{(u,v)} \ge 1-\frac{4}{\sqrt{\beta}}$.
\end{enumerate}
\end{claim}
\begin{proof}
We verify (i) and (ii).
\begin{enumerate}
    \item Suppose $f_{(u,v)} > 0$, then $(u,v)\in H$, so $\dH(u)+\dH(v) \in \{\beta-1,\beta\}$.
    This means that $0\le r_v+r_u\le 1$. If both $r_v$ and $r_u$ are non-negative,
    we have that $x_u + x_v \le 2(\frac{1}{2}+\frac{1^2}{\beta}) \le 1+\frac{1}{\beta}$.
    Now, without loss of generality $r_u < 0 \le r_v$. In case
    $r_u^2 \ge \beta$, we know $x_u = 0$ so $x_u+x_v \le 1$. In the case when
    $r_u^2 < \beta$, we know $x_u = \frac{1}{2}+\frac{r_u^2}{\beta}$ and
    $x_v \le \frac{1}{2}+\frac{r_v^2}{\beta}$. Since $r_u+r_v\le 1$, we known that
    $r_v \le |r_u|+1$. Concluding:
    \begin{equation*}
    x_v+x_u \le 1+\frac{(|r_u|+1)^2-r_u^2}{\beta} = 1+\frac{2|r_u|+1}{\beta} < 1+\frac{2\sqrt{\beta}+1}{\beta} = 1+\frac{2}{\sqrt{\beta}}+\frac{1}{\beta}.
    \end{equation*}
    \item Suppose $x_{v} > 0$. Hence $r_v > -\sqrt{\beta}$ (else $x_v=0$), that is $\dH(v)> \frac{\beta-1}{2}-\sqrt{\beta}$. For any incident edge $(u,v)\in H$, we must have $\dH(v)+\dH(u) \le \beta$, so $\dH(u) \le \frac{\beta-1}{2}+(1+\sqrt{\beta})$. Now, we see that we assign weight at least $1/(\frac{\beta-1}{2}+(1+\sqrt{\beta}))$ to the edge $(u,v)$ in $f$. Since this holds for all the $\dH(v) > \frac{\beta-1}{2}-\sqrt{\beta}$ incident edges we know that $v$ will in total 
    receive, from the fractional matching $f$, at least:
    \begin{equation*}
    \sum_{u:(u,v)\in H} f_{(u,v)} \ge \frac{\tfrac{\beta-1}{2}-\sqrt{\beta}}{\tfrac{\beta-1}{2}+1+\sqrt{\beta}}
 = 1-\frac{4\sqrt{\beta}+2}{(\sqrt{\beta}+1)^2}
    \ge 1-\frac{4}{\sqrt{\beta}}.\qedhere{}
    \end{equation*}
\end{enumerate}
\end{proof}

\begin{proof}[Proof of \cref{thm:explicit-fractional-matching}.]
By the above claims and approximate complimentary slackness (see \cref{lem:approx-cs} in \cref{appendix:proofs}) we know that $(1-\frac{4}{\sqrt{\beta}})|x| \le (1+\frac{2}{\sqrt{\beta}}+\frac{1}{\beta})|f|$. Since $(1-\frac{4}{\sqrt{\beta}}) / (1+\frac{2}{\sqrt{\beta}}+\frac{1}{\beta}) > 1-\frac{6}{\sqrt{\beta}}$, we get that $|f| \ge (1-\frac{6}{\sqrt{\beta}})|x| \ge (1-\eps) |x| \ge (1-\eps)\mu^*(G)$ whenever $\beta \ge 36/\eps^2$.
\end{proof}

\subsection{Lower Bound} \label{sec:lower-bound}

In this section we show that \cref{thm:explicit-fractional-matching} is tight up to a constant, i.e. that one must set $\beta = \Theta(1/{\eps^2})$ in order to guarantee that a weighted EDCS preserves a $(1-\eps)$-fraction of the matching. This might be a bit surprising, considering that for the \emph{unweighted} EDCS, it is known that $\beta = \Theta(1/\eps)$ suffices (to preserve a $(\frac{2}{3}-\eps)$-approximation to the matching \cite{SOheilTIghtAnalysis}).

\begin{figure}[!ht]
\begin{center}
\includegraphics[width=.95\linewidth]{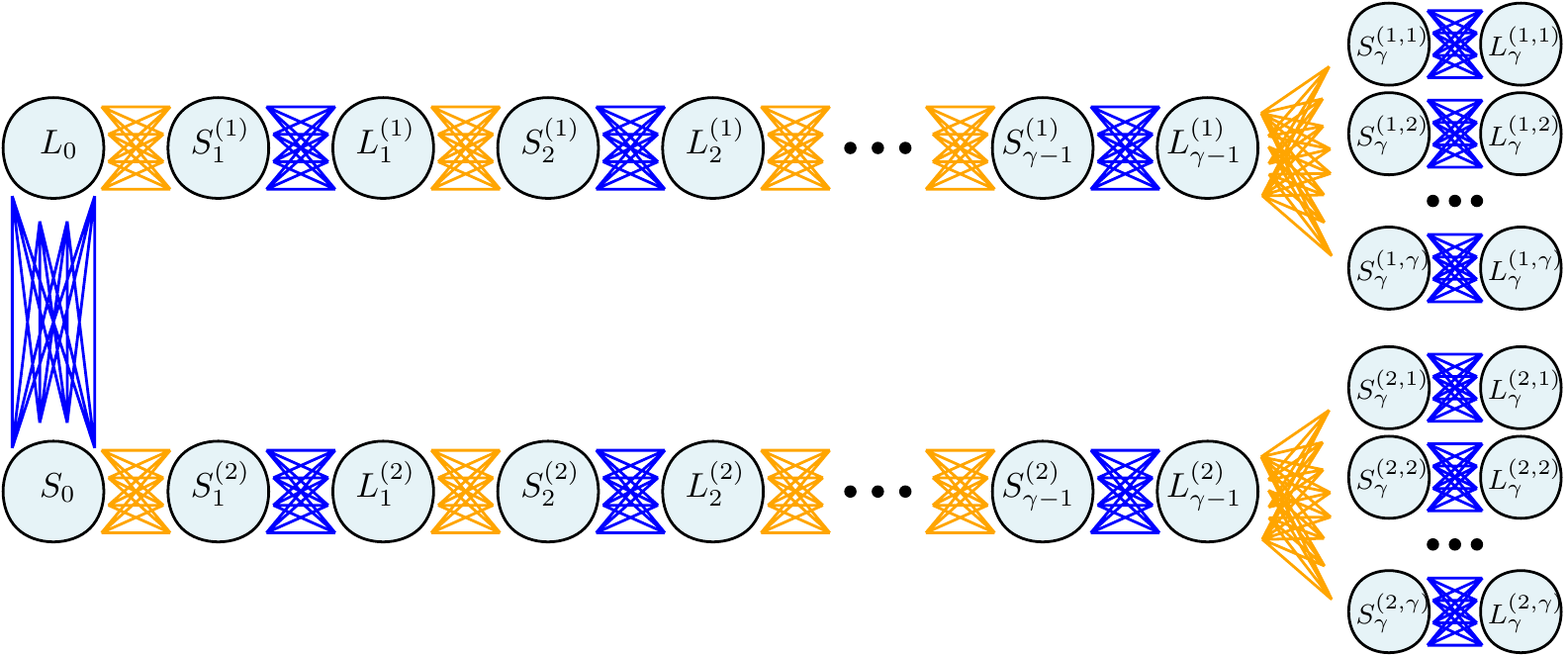}
\end{center}
\caption{The lower bound construction. The blue edges are part of $H$, while the yellow are not. We have $\gamma = \sqrt{\beta/2}$, and each set $S_i$, $L_i$ indicates an independent set of vertices of size $|S_i| = \gamma^2+i$ and $|L_i| = \gamma^2-1$ (so in $H$ they have degrees $\frac{\beta}{2}-i$ and $\frac{\beta}{2}+i$ respectively). The maximum matching in $H$ matches all vertices in $L_i$ to the corresponding $S_i$. The maximum matching in $G$ however, matches $L_i$ to $S_{i+1}$ (and $S_0$ to $S^{(2)}_1$), in addition to $L_{\gamma}$ which can also be matched to $S_{\gamma}$.}
\label{fig:lower-bound}
\end{figure}

\begin{theorem}\label{thm:lower-bound}
For any $\beta\ge 2$, there exists a
(bipartite) graph $G$ together with a weighted EDCS $H$ for which $\mu(H) = (1-\Theta(1/\sqrt{\beta}))\mu(G)$. 
\end{theorem}
We show our construction in \cref{fig:lower-bound}, and also describe it here formally in words.
For simplicity, we will assume that $\beta = 2\gamma^2$ for some integer $\gamma$ (but it is not difficult to adapt the proof for when $\beta$ is not twice a square).
In our construction, each edge appears at most once in $H$,  and all edges $e\in H$ have $\dH(e) = \beta$; all edges $e\in E\setminus H$ have $\deg_H(e) = \beta-1$.

Define the gadget $G_i = (S_i, L_i, E_i)$ to be a complete bipartite graph in which $|S_i| = \gamma^2+i$ and $|L_i| = \gamma^2 - i$ ($S$ is for vertices with \emph{small} degree, and $L$ for vertices with \emph{large} degree).
The subgraph $H$ will consist of many of these gadgets $G_i$, so we start by noting a few properties about them.
Firstly, vertices in $L_i$ have degree $\frac{\beta}{2}+i$ while those in $S_i$ have degree $\frac{\beta}{2}-i$. This means that any edge in $(u,v)\in E_i$ has degree exactly $\deg_{G_i}(u)+\deg_{G_i}(v) = \beta$.
We also note that the maximum matching inside $G_i$ is of size $|L_i| = \gamma^2-i$.

\subparagraph{Subgraph $H$.}
We begin by describing how the weighted EDCS $H$ looks like, and later we will define what additional edges are also in the full graph $G$.
The subgraph $H$ will exactly consist of:
\begin{itemize}
\item One copy of $G_0$.
\item Two copies each of $G_1, G_2, \ldots, G_{\gamma-1}$. Call the copies $G^{(1)}_i$ and $G^{(2)}_i$.
\item $2\gamma$ copies of $G_\gamma$. Call the copies $G^{(1,j)}_{\gamma}$ and $G^{(2,j)}_{\gamma}$ for $j = 1,2,\ldots,\gamma$.
\end{itemize}
\begin{claim}\label{clm:lb-matching-h}
$\mu(H) = 4\gamma^3-4\gamma^2+\gamma$.
\end{claim}
\begin{proof}
Since the maximum matching size in $G_i$ is $|L_i| = \gamma^2-i$ we get:
\begin{align*}
\mu(H) &=
|L_0| + 2\gamma|L_{\gamma}| + 2(|L_1|+|L_2|+\cdots+|L_{\gamma-1}|)\\ &=
\gamma^2 +  2\gamma(\gamma^2-\gamma) + 2\sum_{i=1}^{\gamma-1}(\gamma^2-i) \\ &=
4\gamma^3-4\gamma^2+\gamma\qedhere{}
\end{align*}
\end{proof}

\subparagraph{Full graph $G$.}
Now we describe the additional edges which are part of $G$ but not already in $H$ (see also \cref{fig:lower-bound}):
\begin{itemize}
\item For $k\in \{1,2\}$, we connect $G^{(k)}_1, G^{(k)}_2, \ldots, G^{(k)}_{\gamma-1}$ in a chain as follows: every pair $(u,v)$ with $u\in L^{(k)}_{i}$ and $v\in S^{(k)}_{i+1}$ is an edge (so that the induced subgraph on these two sets of vertices forms a complete bipartite graph). 
\item At the end of these two chains, we connect all the gadgets  $G^{(k,j)}_{\gamma}$ as follows:
every pair $(u,v)$ with $u\in L^{(k)}_{\gamma-1}$ and $v\in S^{(k,j)}_{\gamma}$ for some $j$, is an edge.
\item Finally we connect these two chains using $G_0 = (S_0,L_0,E_0)$ as follows:
every pair $(u,v)$ with $u\in L_{0}$ and $v\in S^{(1)}_{1}$ is an edge; and every
pair $(u,v)$ with $u\in S_{0}$ and $v\in S^{(2)}_{1}$ is an edge.
\end{itemize}
We note that $G$ is bipartite and all above edges have degree exactly $\dH(u)+\dH(v) = \beta-1$, so indeed $H$ is a weighted EDCS of $G$.
\begin{claim}\label{clm:lb-matching-g}
$\mu(G) \ge 4\gamma^3-3\gamma^2+\gamma$.
\end{claim}
\begin{proof}
We argue that a matching of this size exists in $G$. In fact the only edges of $H$ we will use as part of this matching are those in the gadgets $G^{(k,j)}_{\gamma}$.
\begin{itemize}
\item We pick a matching between $L^{(k)}_{i}$ and $S^{(k)}_{i+1}$ of size $|L^{(k)}_i| = \gamma-i$ for all $k\in \{1,2\}$ and $i = 1,2,\ldots,\gamma-2$.
\item In $G^{(k,j)}_\gamma$ we pick a matching of size $|L^{(k,j)}_\gamma| = \gamma^2-\gamma$. Note that exactly $2\gamma$ vertices in $S^{(k,j)}_\gamma$ are left unmatched.
\item Denote by $U^{(k)}$ the set of unmatched vertices in $S^{(k,1)}_\gamma, S^{(k,2)}_\gamma, \ldots, S^{(k,\gamma)}_\gamma$, for $k\in \{1,2\}$.
Note that $|U^{(k)}| = 2\gamma^2$ and that $(L^{(k)}_{\gamma-1},U^{(k)})$  forms a complete bipartite graph, so we pick a matching of size $|L^{(k)}_{\gamma-1}| = \gamma^2-\gamma+1$ from there.
\item Finally we pick matchings between $L_{0}$ and $S^{(1)}_{1}$ (respectively $S_0$ and $S^{(2)}_1$) of size $|L_0| = \gamma$ (respectively $|S_0|=\gamma$).
\end{itemize}
In total we see that the above matching is exactly $|S_0| = \gamma$ larger than in \cref{clm:lb-matching-h}, which concludes the proof of the claim.
\end{proof}

\subparagraph{Approximation ratio.} To conclude the proof of \cref{thm:lower-bound},
we see that $\mu(G)-\mu(H) = \gamma^2 \ge \frac{1}{4\gamma} \mu(G)$ whenever $\gamma \ge 1$. Hence $H$ preserves at most a $(1-\frac{1}{4\gamma}) = (1-\frac{1}{2\sqrt{2\beta}})$ fraction of the maximum matching of $G$.

\bibliography{references.bib}

\appendix
\section{Omitted Proofs}
\label{appendix:proofs}

\paragraph*{EDCS properties.}
\clmdegrees*
\begin{proof}
If a vertex $u$ has $\dH(u) > \beta$, then any incident edge $(u,v)\in H$ is overfull: $\dH(u)+\dH(v) > \beta$, leading to a contradiction. Hence the maximum degree is at most $\beta$. Now we construct a fractional matching by assigning a weight of $1/\beta$ to every edge in $H$ (so an edge appearing with multiplicity $\phi$ in $H$ gets weight $\phi/\beta$). Clearly this is a feasible fractional matching of $G$, since no vertex is overloaded. On the other hand, the size of this fractional matching is $|H|/\beta$, implying that
$|H| \le \beta \mu^*(G)$.
\end{proof}

\clmupdates*

\begin{proof}
We use a potential function argument. Define
\begin{equation*}
\Phi(H) := |H|(2\beta-1) - \sum_{(u,v)\in H} (\dH(v)+\dH(u))
= \sum_{(u,v)\in H} (2\beta-1 - \dH(u)-\dH(v))
\end{equation*}
We note that if an edge $(u,v)$ appears multiple times in $H$, it appears multiple times in the above sum as well. We first note that $\Phi(\emptyset)=0$ and $\Phi(H) \le \beta|H| \le \beta^2\mu^*(G)$, by \cref{clm:degrees} and since $(2\beta-1-\dH(u)-\dH(v))\le \beta$ when $H$ is a valid EDCS. Now we verify that updates to $H$ increase the potential by at least 1:
\begin{itemize}
\item Insertion of an underfull edge $(u,v)\in E$.

That is $\dH(u)+\dH(v)\le \beta-2$ before adding the edge.
The term $|H|(2\beta-1)$ will increase by $2\beta-1$. $\sum_{(u,v)\in H} (\dH(v)+\dH(u))$ will increase by at most $2\beta-2$, since one term of value $\dH(v)+\dH(v)\le \beta-2+2$ (the $+2$ comes from $\dH(v)$ and $\dH(u)$ increasing by one when we add $(u,v)$  to $H$) is added, and at most $\beta-2$ other terms decrease in value by one.

\item Deletion of an overfull edge $(u,v)\in H$.

That is $\dH(u)+\dH(v)\ge \beta+1$ before removing the edge.
The term $|H|(2\beta-1)$ will decrease by $2\beta-1$. $\sum_{(u,v)\in H} (\dH(v)+\dH(u))$ will decrease by at least $2\beta$, since one term of value $\dH(v)+\dH(v)\ge \beta+1-2$ (the $-2$ comes from $\dH(v)$ and $\dH(u)$ decreasing when we remove $(u,v)$) is erased, and at least $\beta+1$ other terms increase in value by one. \qedhere
\end{itemize}
\end{proof}

\clmupdatesvertex*
\begin{proof}
We continue the potential function argument from the proof of \cref{clm:updates} above. When we delete, from $H$, all edges incident to some vertex $u$, we know that we deleted at most $\beta$ many edges from $H$ (as the degree of this vertex was at most $\beta$). For each such incident edge $(u,v)$, we bound how much its deletion could have decreased the potential function. The $|H|(2\beta-1)$ term decreased by exactly $2\beta-1$, and the $-\sum_{(u,v)\in H}(\dH(u)+\dH(v))$ term can only increase. So the total decrease in potential, over all up to $\beta$ incident edges which were deleted, is at most $2\beta^2-\beta$.

Since we can only delete up to $n$ vertices in total, and the potential is always bounded by $\beta^2 \mu^*(G)\le \beta^2 n$, it follows that the total increase in the potential function, over the run of the algorithm, is at most $3\beta^2 n - n\beta$ (and thus this many updates to $H$ from insertions/deletions of underfull/overfull edges). In total we deleted at most $n\beta$ edges in $H$ incident to deleted vertices, so the total number of updates to $H$ is thus bounded by $3\beta^2 n -\beta n + \beta n = 3\beta^2 n$.
\end{proof}

\paragraph*{Approximate Complimentary Slackness.}
\begin{lemma}
\label{lem:approx-cs}
Suppose we have the primal linear program $\max\{c^T x : Ax\le c, x\ge 0\}$, and its dual $\min\{c^T y : A^T y \ge b, y\ge 0\}$. We say that feasible primal solution $x$ and dual solution $y$ satisfy $(\alpha, \gamma)$-approximate complementary slackness (for $\alpha \le 1 \le \gamma$) if: (i) if $x_i = 0$ then $(A^T)_i y \le \gamma b_i$, and (ii) if $y_j = 0$ then $(A)_j x \ge \alpha$. When this is the case, then $\gamma b^T y \le \alpha c^T x$ (i.e.\ $x$ and $y$ are $\frac{\gamma}{\alpha}$-approximate optimal).
\end{lemma}
\begin{proof}
We see that $\alpha c^T x - \gamma b^T y = x^T(\alpha b - A^Ty) + y^T(Ax-\gamma b)$. Now either $x_i = 0$ or $(\alpha b - A^T y)_i \le 0$; and
either $y_j = 0$ or $(Ax - \gamma b)_j \le 0$. Hence $\alpha c^T x - \gamma b^T y \ge 0$.
\end{proof}

\end{document}